\documentstyle[aps,prd,t1enc,amsmath,mathrsfs,cite]{revtex}

\newcommand{\hs}{\,-\,}

\begin{document}

\title{Plane fronted parallel waves in a warm two\hs component plasma} 

\author{Gert Brodin$^1$\footnote{E\hs mail:
    gert.brodin@physics.umu.se},
  Mattias Marklund$^2$\footnote{E\hs mail:
    mattias.marklund@elmagn.chalmers.se}
  and Peter K.\ S.\
  Dunsby$^3$\footnote{E\hs mail: peter@vishnu.mth.uct.ac.za}}

\address{$1$ Department of Plasma Physics,
  Ume{\aa} University, SE--901 87 Ume{\aa}, Sweden}
\address{$2$ Department of Electromagnetics, Chalmers University of
  Technology, SE--412 96 G\"oteborg, Sweden}
\address{$3$ Department of Mathematics and Applied Mathematics,
  University of Cape Town, Rondebosch 7701, South Africa}

\date{\today}

\maketitle

\begin{abstract}
A general system of equations is derived, using the 1+3 orthonormal tetrad 
formalism, describing the influence of
a plane\hs fronted\hs  parallel gravitational wave on a warm relativistic
two\hs component plasma. We focus our attention on phenomena that
are induced by terms that are higher order in the gravitational wave
amplitude. In particular, it is shown that parametric excitations of
ion\hs acoustic waves takes place, due to these higher order gravitational
non\hs linearities. The implications of the results are discussed.
\end{abstract}

\draft

\section{Introduction}

There have been many theoretical investigations 
on the scattering of electromagnetic waves off gravitational fields, 
using linearized gravitational wave theory (see e.g. \cite{bi:Collection}), 
which is the relevant regime for gravitational wave detectors, 
or, in general, for distances far away from the gravitational 
wave source. 
Much of previous research has been directed at studying effects on 
vacuum electromagnetic fields, but there has also been some work 
where the effects of plasmas have been taken into account (see e.g.\ 
\cite{bi:Marklund-Brodin-Dunsby,bi:rest,bi:Brodin-Marklund-Dunsby,%
Servin} and references therein ). 

In a recent paper \cite{bi:Brodin-Marklund-Dunsby} we 
have taken another approach, starting with an exact 
gravitational wave solution, but focusing on a weak amplitude 
(but still non\hs linear) approximation, and studying the effects 
induced by gravitational waves in a plasma. As was shown in 
\cite{bi:Brodin-Marklund-Dunsby}, such interactions give 
rise to qualitatively new phenomena that are absent 
in linearized theory. However, the previous work was limited to cold 
plasmas, and thus did not allow interaction with low\hs frequency 
longitudinal waves (although for example Ref.\ \cite{Servin} treats the 
generation of low\hs frequency plasma modes by {\em linear}
gravitational waves). 
As will be shown, such waves are more likely to be 
excited by gravitational waves,\footnote{As compared to the high\hs 
  frequency excitation process considered in Ref.\ 
  \cite{bi:Brodin-Marklund-Dunsby}.} due to the frequency matching 
condition.

We start by generalizing previous equations for a cold two\hs component 
plasma to include thermal effects, using the 1+3 orthonormal frame 
formalism. Within this mathematical framework, the governing plasma equations 
can be written in a simple 3\hs dimensional form. They consist of 
Maxwell's equations with additional charge and current densities 
characterizing the gravitational effects and a set of fluid equations 
for a warm plasma. To facilitate the
analysis of the non\hs linear interaction between a plasma and a
gravitational wave, we make use of the \emph{plane fronted parallel} 
(pp) wave solution of Einstein's field equations. 
We show that for parallel propagation, excitation of ion\hs acoustic waves 
can only occur if effects second order in the gravitational wave amplitude
are included. The growth rate of the second order instability is 
determined, and the threshold value for excitation is estimated.  
Some applications to astrophysics and cosmology are discussed and our 
results are summarized in the last section of the paper.

\section{Preliminaries}

The energy\hs momentum tensor for each particle species is assumed to
take the form of a perfect fluid,  $T^{ab} = (\mu + p)V^aV^b + pg^{ab}$, 
where $\mu$ is the energy density and $p$ the pressure of each fluid.
Splitting the energy\hs momentum tensor in time- and spacelike parts,
using an orthonormal frame $\{{\boldsymbol{e}}_a, a = 0, ..., 4\}$,
the particle and momentum conservation equations take the form 
\cite{Marklund} 
\begin{mathletters}\label{conservation}
  \begin{eqnarray}
    && {\boldsymbol{e}}_0(\gamma n) 
    + {\boldsymbol{\nabla\cdot}}(\gamma
    n{\boldsymbol{v}}) = -\gamma n\left( \Gamma^{\alpha}\!_{0\alpha}
    + \Gamma^{\alpha}\!_{00}v_{\alpha} +
    \Gamma^{\alpha}\!_{\beta\alpha}v^{\beta} \right) \ , \\
    && (\mu + p)\left( {\boldsymbol{e}}_0 + {\boldsymbol{v\cdot\nabla}}
    \right)\gamma{\boldsymbol{v}} + \gamma^{-1}{\boldsymbol{\nabla}}p 
    + \gamma{\boldsymbol{v}}\left( {\boldsymbol{e}}_0 
    + {\boldsymbol{v\cdot\nabla}}
    \right)p \nonumber \\
    && \qquad= qn({\boldsymbol{E}}
    + {\boldsymbol{v \times B}})
    - \gamma(\mu + p)\left[ \Gamma^{\alpha}\!_{00} +
    \left( \Gamma^{\alpha}\!_{0\beta} + \Gamma^{\alpha}\!_{\beta 0}
    \right)v^{\beta} +
    \Gamma^{\alpha}\!_{\beta\gamma}v^{\beta}v^{\gamma}
    \right]{\boldsymbol{e}}_{\alpha}  \ ,
  \end{eqnarray}
\end{mathletters}
generalizing the results presented in \cite{bi:Brodin-Marklund-Dunsby}
to the case of a warm plasma.
Here $\Gamma^a\!_{bc}$ are the Ricci rotation coefficients with 
respect to the ONF $\{{\boldsymbol{e}}_a\}$. We have introduced the
notation ${\boldsymbol{E}} \equiv (E^{\alpha}) = (E^1,
E^2, E^3)$ etc., ${\boldsymbol{\nabla}} \equiv ({\boldsymbol{e}}_1,
{\boldsymbol{e}}_2, {\boldsymbol{e}}_3)$. The fluid is assumed to have 
a four\hs velocity $(\gamma, \gamma{\boldsymbol{v}})$, relative to the 
orthonormal frame $\{{\boldsymbol{e}}_a, a = 0, ..., 4\}$, where 
$\gamma \equiv (1 - v^2)^{-1/2}$. Furthermore, 
$\rho \equiv q\gamma n$ is the charge density.

We follow the approach presented in Refs.\ \cite{bi:Marklund-Brodin-Dunsby,%
bi:Brodin-Marklund-Dunsby} for handling gravitational effects in 
Maxwell's equations. With this, these equations was presented in their 
generality in \cite{bi:Marklund-Brodin-Dunsby,%
bi:Brodin-Marklund-Dunsby}.
In order to address the issue of how strong gravitational
radiation may be involved in generation of EM waves, we look at
the pp\hs waves
(for a discussion of this solution, see \cite{bi:pp-wave}), 
in the special case of a linearly polarized plane wave the spacetime
metric takes the following form
\begin{equation}\label{eq:pp}
  {\rm d}s^2 = -{\rm d}t^2 + a(u)^2\,{\rm d}x^2 + b(u)^2\,{\rm d}y^2 +
  {\rm d}z^2  \ ,
\end{equation}
where $u = z - t$, and $a$ and $b$ satisfy $ab_{uu} + a_{uu}b =
0$, and the subscript $u$ denotes a derivative with respect to retarded
time. Note that we have chosen a vacuum geometry, i.e. we have 
omitted the influence of the plasma on the metric.

In order to make physical interpretations simple, we introduce the
canonical Lorentz frame
\begin{equation}\label{frame}
  {\boldsymbol{e}}_0 = \partial_t \ , \ {\boldsymbol{e}}_1 =
  a^{-1}\partial_x \ , \
  {\boldsymbol{e}}_2 = b^{-1}\partial_y \ , \ {\boldsymbol{e}}_3 =
  \partial_z \ .
\end{equation}
With this choice, the gravitational effects in Maxwell's equations
are
\begin{mathletters}\label{bongo}
\begin{eqnarray}
  \rho_{\!_E} &=& -(\ln ab)_uE^3 \ , \label{eq:e-charge} \\
  \rho_{\!_B} &=& -(\ln ab)_uB^3 \ , \label{eq:b-charge} \\
  {\boldsymbol{j}}_{\!_E} &=& -(\ln b)_u(E^1 -
  B^2)\,{\boldsymbol{e}}_1 - (\ln a)_u(E^2 + B^1)\,{\boldsymbol{e}}_2
  - (\ln ab)_uE^3\,{\boldsymbol{e}}_3   \ ,
  \label{eq:e-current} \\
  {\boldsymbol{j}}_{\!_B} &=& -(\ln b)_u(E^2 +
     B^1)\,{\boldsymbol{e}}_1 + (\ln a)_u(E^1 -
     B^2)\,{\boldsymbol{e}}_2 - (\ln ab)_uB^3\,{\boldsymbol{e}}_3  \ ,
      \label{eq:b-current}
\end{eqnarray}
\end{mathletters}
which enters Maxwell's equations as
\begin{mathletters} \label{Maxwell}
\begin{eqnarray}
  {\boldsymbol{\nabla\cdot E}} &=& \rho_{\!_E} + \rho_{\rm ch}
  \label{eq:max1} \ , \\
  {\boldsymbol{\nabla\cdot B}} &=& \rho_{\!_B} \label{eq:max2}\ , \\
  \dot{\boldsymbol{E}} - {\boldsymbol{\nabla\times B}}
     &=& -{\boldsymbol{j}}_{\!_E} - {\boldsymbol{j}} \label{eq:max3}
     \ , \\
  \dot{\boldsymbol{B}} + {\boldsymbol{\nabla\times E}}
     &=& -{\boldsymbol{j}}_{\!_B}  \label{eq:max4} \ ,
\end{eqnarray}
\end{mathletters} 
where ${\boldsymbol{j}} \equiv \sum_{\rm p.s.}q\gamma n{\boldsymbol{v}}$ 
is the current density, and the overdot stands for
time derivative.

With respect to the frame choice (\ref{frame}), 
the conservation equations (\ref{conservation}) lead to the  
following fluid equations:
\begin{mathletters}\label{fluideqs}
\begin{eqnarray}
  && \frac{\partial}{\partial t}(\gamma n)
    + {\boldsymbol{\nabla\cdot}}(\gamma n{\boldsymbol{v}})
  = \gamma n(\ln ab)_u(1 - v_{\parallel})
    \ , \label{energy} \\
  && (\mu + p)\left(\frac{\partial}{\partial t} + {\boldsymbol{v\cdot\nabla}}
  \right) \gamma{\boldsymbol{v}} + \gamma^{-1}{\boldsymbol{\nabla}}p 
  + \gamma{\boldsymbol{v}}\left(\frac{\partial}{\partial t} 
    + {\boldsymbol{v\cdot\nabla}}\right)p \nonumber \\
  && \qquad = \rho_{\rm ch}({\boldsymbol{E}} + {\boldsymbol{v\times B}})
  + \gamma(\mu + p)\left[ (\ln a)_uv_1{\boldsymbol{e}}_1 + (\ln
b)_uv_2{\boldsymbol{e}}_2 \right](1 - v_{\parallel})
  + \gamma(\mu + p)\left[ (\ln a)_uv_1^2 +(\ln b)_uv_2^2
    \right]{\boldsymbol{e}}_3  \ , \label{momentum}
\end{eqnarray}
\end{mathletters}
where $v_{\parallel} \equiv v_3$ is the velocity parallel to the
gravitational wave propagation direction. These equations should be
satisfied for each particle species. 

\section{Longitudinal wave excitations}

The terms in equations (\ref{bongo})--(\ref{fluideqs}) which generate 
effects in the direction of propagation of the gravitational wave, 
i.e.\ the longitudinal terms, are second order in the gravitational 
wave amplitude. As was previously shown in 
\cite{bi:Brodin-Marklund-Dunsby}, these second order terms can 
give rise to qualitatively new phenomena compared to the linear regime. 
However, for the process described in \cite{bi:Brodin-Marklund-Dunsby}, 
the gravitational wave frequency must be equal to the local plasma 
frequency, which puts rather severe constraints on the possible sources 
for the gravitational waves.
Here we will focus on the excitation of longitudinal waves with much 
lower frequencies, specifically ion\hs acoustic waves. For simplicity, 
the equation of state is assumed to be an non\hs relativistic 
isothermal,\footnote{A more 
general temperature\hs to\hs density profile can 
easily be incorporated into the  calculations.} 
i.e.\ $p = k_BTn$, $k_B$ being the Boltzmann constant and $T$ the 
temperature.  

As a consequence of the low temperature, the inertial mass term  
$\mu + p \approx mn$ and it is this approximation that will be used 
from now on. Furthermore, we consider the regime $\ln(ab) \ll 1$. 
The background is homogeneous in both fluid species, and thus we can 
take the zeroth order electromagnetic fields to be zero. 
Neglecting the effect of charge separation, we obtain the first order 
solutions to Eqs.\ (\ref{fluideqs})
\begin{equation}\label{equilibrium}
  n_{\rm gw}(u) = -n_0\frac{\ln(ab)}{1 - v_{\rm th}^2} \ , \quad 
  v_{\rm gw}(u) = -v_{\rm th}^2\frac{\ln(ab)}{1 - v_{\rm th}^2} \ ,
\end{equation}
where $n_0$ is the background number density and $v_{\rm th} 
\equiv \sqrt{k_BT/m}$ the thermal velocity. The above equations 
hold for both particle species so that the density perturbations 
are in general different for the two fluids.\footnote{In principle the 
  electric field due to the possible charge separation should be included
  in Eq.\ (\ref{equilibrium}). However, since this effect is proportional
  to both $(\ln ab)_u$ and $v_{\rm th}^2$, it will not influence our main 
  result (\ref{growth}).}

With the above calculation as a prerequisite, we consider the 
stability of these solutions, i.e.\ we make the ans\"atze 
\begin{equation}
  n(t,z) = n_0 + n_{\rm gw}(u) + \hat{n}(t,z) \ , \quad 
  v(t,z) = v_{\rm gw}(u) + \hat{v}(t,z) \ , \quad 
  E(t,z) = \hat{E}(t,z) \ .
\end{equation}
We linearize in $\hat{n}$, $\hat{v}$ and $\hat{E}$, and keep terms up to order
$(\ln ab)_u\times$($\hat{n}$, $\hat{v}$ or $\hat{E}$).
The effect of terms quadratic or higher order in $(\ln ab)_u$ is to modify 
the equilibrium solution (\ref{equilibrium}) by a factor $\sim 1 + (\ln ab)$. 
Thus we neglect the corresponding influence on the wave excitation. 

Using the above perturbation scheme, we can combine equations (\ref{energy}) 
and (\ref{momentum}) to obtain
\begin{equation}\label{density}
  {\mathscr{D}}\hat{n} = \frac{\partial S}{\partial t} 
    - \frac{n_0q}{m}\frac{\partial\hat{E}}{\partial z} \ , 
\end{equation}
where 
\begin{equation}
  {\mathscr{D}} \equiv 
  \frac{\partial^2}{\partial t^2} - v_{{\rm th}}\frac{\partial^2}{\partial z^2}
\end{equation} 
and
\begin{equation}
  S \equiv (n_0\hat{v} - \hat{n})\frac{\partial}{\partial t}(\ln ab) 
    - \frac{\partial}{\partial z}(n_{\rm gw}\hat{v} + \hat{n}v_{\rm gw}) \ . 
\end{equation}
Note that ${\mathscr{D}}$ and $S$ in general are different for different 
particle species.

Acting with ${\mathscr{D}}_1{\mathscr{D}}_2$ on Eq.\ (\ref{eq:max1}), we 
obtain 
\begin{equation}\label{waveeq}
  \left( {\mathscr{D}}_1{\mathscr{D}}_2  
    + \omega_{\rm p1}^2{\mathscr{D}}_2 + \omega_{\rm p2}^2{\mathscr{D}}_1 \right)%
  \frac{\partial\hat{E}}{\partial z}
  = q_1{\mathscr{D}}_2\frac{\partial S_1}{\partial t} +
    q_2{\mathscr{D}}_1\frac{\partial S_2}{\partial t} +
    {\mathscr{D}}_1{\mathscr{D}}_2%
    \left[ \hat{E}\frac{\partial}{\partial t}(\ln ab) \right] \ ,
\end{equation}
where the indices $1$ and $2$ denotes the two fluid species, and 
$\omega_{\rm p1,2} \equiv [q_{1,2}^2n_0/m_{1,2}]^{1/2}$ is the 
plasma frequency of each particle species. 
Equation (\ref{waveeq}) holds for {\em any} two\hs fluid plasma with low 
temperature in both species, and can describe excitation of both high\hs 
frequency and low\hs frequency waves.
A process 
involving high\hs frequency waves in a cold, one\hs component plasma, 
was investigated in \cite{bi:Brodin-Marklund-Dunsby}. 
However, in most cases the gravitational wave frequency is lower than
the electron plasma frequency. Thus, from now on we will focus on three\hs 
wave excitation of low\hs frequency modes (i.e. all frequencies $\ll 
\omega_{\rm p,e}$) in an electron\hs ion plasma. Note that the gravitational 
wave acts as our pump wave. We are interested in the case of a periodic source
of our pump wave, and following \cite{bi:Brodin-Marklund-Dunsby} 
we obtain (to second order in an expansion in the amplitude $h$)
\begin{equation} 
  (\ln ab) = \frac12h^2\exp[2i\omega_{\rm gw}(z - t)] + {\rm c.c.} \ ,
\end{equation}
where c.c.\ denotes the complex conjugate. 

The electric field perturbation is assumed to have the form 
\begin{equation}
  \hat{E} = E_+(t)\exp[i(k_+z - \omega_+t)] 
          + E_-(t)\exp[i(k_-z - \omega_-t)] + {\rm c.c.} \ ,
\end{equation}
and the analogous expressions is assumend to hold for $\hat{n}$ and 
$\hat{v}$. The time variations of  the amplitudes are slow, as 
compared to $\omega_\pm$. The wave numbers and frequencies satisfy 
the matching conditions
\begin{equation}\label{matching}
  k_+ + k_- = 2\omega_{\rm gw} \ , \quad 
  \omega_+ + \omega_- = 2\omega_{\rm gw} \ .
\end{equation}
However, for low\hs frequency waves, the phase velocities are
generally much smaller than unity, which means that equation (\ref{matching})
can be approximated by
\begin{equation}\label{matching2}
  k_+ = -k_- \equiv k \ , \quad \omega_+ = \omega_- = \omega_{\rm gw} \ .
\end{equation}
To be consistent, we should simplify the right hand side of Eq.\ 
(\ref{waveeq}) as far as possible, using $v_{\rm th} \ll 1$, which, 
for example, implies $S = -\hat{n}\partial(\ln ab)/\partial t$.
Inserting the ans\"atze for $\hat{n}$, $\hat{v}$ and $\hat{E}$ in Eq.\ 
(\ref{waveeq}), eliminating the density variations using Eq.\ (\ref{density}), 
and applying the approximate matching conditions 
(\ref{matching2}), results in a growth of the amplitudes 
$\hat{n}, \hat{v}, \hat{E} \propto \exp(\Gamma t)$, with the growth rate
\begin{eqnarray}
  \Gamma &=& 2\left( \frac{\partial D(\omega_{\rm gw},k)}%
  {\partial\omega_{\rm gw}} \right)^{-1} h^2\left[ \frac{\omega_{\rm gw}}{k} 
    + \omega_{\rm gw}^2%
  \sum_{\rm p.s.}\frac{\omega_{\rm p}^2}{(\omega_{\rm gw}^2 - k^2v_{\rm th})^2}
  \right] \nonumber \\
  && \quad \approx 2\left( \frac{\partial D(\omega_{\rm gw},k)}%
  {\partial\omega_{\rm gw}} \right)^{-1} h^2\omega_{\rm gw}^2%
  \frac{\omega_{\rm p,i}^2}{(\omega_{\rm gw}^2 - k^2v_{\rm th,i})^2}
  \approx h^2\omega_{\rm gw} \ , \label{growth}
\end{eqnarray}
where the sum is over particle species and  
\begin{equation}
  D(\omega,k) = 1 - \sum_{\rm p.s.}%
    \frac{\omega_{\rm p}^2}{\omega^2 - k^2v_{\rm th}^2} \ ,
\end{equation}
and the index i refers to the ions.%
\footnote{The dispersion function $D(\omega,k)$ can 
  be further reduced using $\omega \ll \omega_{\rm p,e}$, e denoting the 
  electrons. Applied to the expression for $\Gamma$,
  we obtain the last approximate equality in Eq.\ (\ref{growth}).} 
The scale length $1/k$ of the excited modes, obtained from 
$D(\omega_{\rm gw},k) = 0$, satisfies
\begin{equation}
  \frac{v_{\rm th,i}}{\omega_{\rm gw}} \leq \frac{1}{k} 
    \leq \frac{c_s}{\omega_{\rm gw}} \ ,
\end{equation}
where $c_s \equiv [v_{\rm th,i}^2 + (m_{\rm e}/m_{\rm i})v_{\rm th,e}^2]^{1/2}$ is 
the ion\hs acoustic velocity.

By including some mechanism of wave damping, a threshold value $h_{\rm thr}$ 
for excitation may be determined. The most suitable regime for excitation is
$T_{\rm e} > T_{\rm i}$ in which case ion Landau damping is not effective. 
Then electron\hs ion collisions is the main disspative mechanism, and 
\begin{equation}
h_{\rm thr} \sim \sqrt{\nu_{\rm e-i}/\omega_{\rm gw}}\;,
\end{equation}
which is similar to the results obtained in \cite{bi:Brodin-Marklund-Dunsby}.
Here $\nu_{\rm e-i}$ is the electron\hs ion collision frequency.

Without working out a detailed theory for the saturation mechanism, 
we note that the wave growth will stop at a level $\hat{v}\leq v_{\rm th,i}$. 
This is because the ion\hs acoustic waves become strongly nonlinear 
at this level, implying efficent energy transport away from the 
originally excited modes.

\section{Summary and discussion}

In the present paper we have generalized previous equations for a cold
plasma in the presence of a strong gravitational wave 
\cite{bi:Brodin-Marklund-Dunsby}, by including thermal effects. As is
well known, thermal effects are important for low frequency plasma phenomena.
Since, typically, the time\hs scales for gravitational waves are long
compared to the time\hs scales of a plasma, this generalization is important
from an applicational point of view. The derived equations provide a framework
for investigating strong gravitational pulse effects in warm multi\hs
component plasmas. It was shown that the equations indeed admit generation
of ion\hs acoustic modes, which are not present in the linearized theory
(cf.\ Ref.\ \cite{Servin}).

Since our effect is of second order in the gravitational wave amplitude,
possible astrophysical applications are most likely to be found close to
extreme events, such as binary mergers. The expression for the growth rate 
is almost identical to that of Ref.\ \cite{bi:Brodin-Marklund-Dunsby}, where
two plasmon decay of a pp\hs wave was considered. However, a fact that  
make the present process more relevant for astrophysical applications is that 
it in principle can occur for arbitrarily low gravitational wave 
frequency $\omega_{\rm gw}$, i.e.\ the plasma frequency is allowed to be 
much larger than $\omega_{\rm gw}$. 

Note that Eq.\ (\ref{equilibrium}) implies that the gravitational wave directly
induces charge separation, provided that the thermal velocities of the
particle species are different (which is in general true for ion\hs electron
plasma). This charge separation could be important in the early universe,
since it could lead to a weak statistically homogeneous and isotropic electric
field. Provided the conductivity of the cosmological plasma remains 
sufficently low, conditions which exist during inflation and 
the subsequent period of reheating, this primordial electric field 
could survive immediate dissipation and could trigger a period of 
cosmological magnetogenesis through its interaction with linear 
gravitational waves \cite{bi:Marklund-Dunsby-Brodin,bi:Christos}. 
In this way large scale cosmological magnetic field could be generated
via physicical processes inherent to plasmas and the geometrical nature
of spacetime, rather than invoking field theoretical arguments such as the 
breaking of conformal invariance of electromagnetism.

\section*{Acknowledgments}

This research was supported by Sida/NRF. G.\ B.\ and M.\ M.\ would like to thank 
the Cosmology Group at the Dept.\ of Mathematics and Applied Mathematics, 
University of Cape Town, for their hospitality.

\end{document}